\documentclass[preprint,12pt]{sn-jnl}

\usepackage[english]{babel}
\usepackage{amsmath,amssymb,amsfonts,amsthm, bm}
\usepackage{dsfont}
\usepackage{mathrsfs}
\usepackage{xcolor}
\usepackage{graphicx}
\usepackage{epstopdf}
\usepackage{array,booktabs,tabularx,longtable,multirow,makecell}
\usepackage{subfigure,caption}
\usepackage{url}
\usepackage{hyperref}
\usepackage{natbib}
\usepackage[ruled]{algorithm2e}
\usepackage{listings}
\usepackage{float}
\usepackage{svg}
\usepackage{pdflscape}
\usepackage{tikz}
\usepackage{pgfplots}
\usepackage{comment}
\usepackage{lastpage}
\usepackage{enumerate}
\usepackage{fancyhdr}
\usepackage{fncylab}
\usepackage[colorinlistoftodos]{todonotes}
\usepackage{algpseudocode, marvosym}
\pgfplotsset{compat=1.18}
\usepgfplotslibrary{groupplots}

\newtheorem{definition}{Definition}

\DeclareMathOperator*{\argmin}{arg\,min}

\SetKwInput{KwInitialize}{Initialize}

\allowdisplaybreaks

\begin{document}

\title{ADMM-based decomposed DNN+RLT Relaxations for Completely Positive Models in Electricity Market Clearing}

\author{Shudian Zhao, Mohammad Reza Karimi Gharigh, Jan Kronqvist, Mohammad Reza Hesamzadeh 

KTH Royal Institute of Technology, Stockholm, Sweden }

\maketitle

\begin{abstract}
\noindent\textbf{Abstract.}
The day-ahead electricity market clearing with nonconvex order types can be formulated as a mixed-integer linear program (MILP), but its LP relaxation may provide weak bounds, and exact solutions can become computationally intractable in large-scale or extended market settings. We study a welfare-maximizing clearing model with elementary hourly orders, block orders with logical acceptance constraints, and flexible hourly orders. Starting from a compact MILP formulation, we derive an equivalent completely positive programming (CPP) reformulation via matrix lifting and propose relaxed CPP variants that further reduce the modeling burden while maintaining strong bounds. We then develop tractable doubly nonnegative (DNN) relaxations, including decomposed formulations that exploit the problem structure by using smaller positive semidefinite matrices. To further strengthen these bounds, we introduce reformulation-linearization technique (RLT) inequalities tailored to the decomposed structure. To tackle the challenge of large-scale DNNs, we design an alternating direction method of multipliers (ADMM) with adaptive penalty updates and rigorous dual lower bounds, enabling certified early termination. Computational experiments on synthetic instances show that the proposed DNN+RLT relaxations substantially tighten LP bounds, while decomposition and first-order methods significantly reduce computational effort.

\vspace{1mm}
\end{abstract} 

\section{Introduction}
Day-ahead electricity market clearing is typically formulated as a welfare-maximization problem over a finite time horizon, subject to market-balance conditions and product-specific feasibility constraints. In practical market designs, however, the presence of nonconvex order types---such as block bids, linked (child--parent) bids, exclusive groups, and flexible hourly products---introduces discrete decisions and logical constraints, leading naturally to large-scale mixed-integer linear programs (MILPs)~\cite{7521190}. Although many operational instances can be solved exactly by MILP technology, strong relaxations remain important for producing informative bounds, supporting scalable algorithms, and enabling richer market models for which an exact solution may become computationally demanding.

Conic relaxations provide stronger bounds than linear relaxations \citep{yokoyama2015optimization}. There have not been many applications of conic relaxations in market-clearing problems, but recent studies on unit commitment problems (problems that share very similar non-convexity and large-scale nature) show the modeling power of semidefinite positive (SDP) relaxations strengthened by RLT inequalities~\citep{fattahi2017conic}. Moreover, Guo, Bodur, and Taylor \cite {guo2024copositive} proposed a completely positive programming (CPP) reformulation for unit commitment in the U.S. market and, based on the strong duality of CPP, established a pricing scheme by calculating dual solutions. 

This paper develops doubly nonnegative relaxation frameworks for market clearing with nonconvex orders. Starting from a compact MILP formulation, we exploit matrix lifting to derive an equivalent completely positive programming (CPP) reformulation. This theoretical result provides a systematic way to encode binary structure and complementarity in a conic framework, which benefits from strong duality, an important feature for pricing mechanisms. Since the complete positive (CP) cone is computationally intractable in general, we next replace the CP cone with tractable outer approximations based on the doubly nonnegative (DNN) cone, leading to semidefinite relaxations that can be substantially tighter than the basic LP relaxation.  The RLT technique provides valid linear inequalities by multiplying existing constraints and then linearizing the resulting product terms. It has long been used to further tighten relaxations for nonconvex optimization problems \cite{sherali1992global}.

To improve scalability, we further exploit sparsity and temporal structure to decompose the lifted matrix into smaller clique matrices linked by consistency constraints. We then strengthen the resulting DNN relaxations with the reformulation-linearization technique (RLT) inequalities tailored to the logical structure of market orders. Finally, for the decomposed primal semidefinite relaxations, we develop an alternating direction method of multipliers (ADMM) scheme with adaptive penalty updates and computable dual lower bounds, allowing certified early termination.
\subsection*{Contributions}
The main contributions of the paper are as follows:
\begin{enumerate}
    \item We derive an equivalent CPP reformulation of the market-clearing MILP using matrix lifting, and introduce relaxed CPP variants that are better suited for scalable implementation.
    \item We propose a decomposition strategy that exploits temporal structure and order-type couplings, replacing a large lifted cone constraint with smaller conic blocks and consistency constraints.
    \item We strengthen the resulting DNN relaxations by adding RLT inequalities derived from box constraints and market-order logic.
    \item We design a practical ADMM method for the decomposed primal semidefinite relaxations, with adaptive penalty updates and rigorous dual lower bounds for certified early termination.
    \item We provide numerical evidence showing that the proposed relaxations substantially tighten LP bounds and that decomposition, together with first-order methods, improves computational scalability.
\end{enumerate}
\subsection*{Outlines}
Section~\ref{sec:market_model} introduces the welfare-maximizing market-clearing model and its MILP formulation. Section~\ref{sec:CPP_reformulation} derives the exact completely positive reformulation. Section~\ref{sec:reformulation_decomposition_relaxation} presents the relaxed and decomposed CPP formulations together with the SDP, DNN, and DNN+RLT relaxations. Section~\ref{sec:ADMM_for_DNN} introduces the ADMM method for the decomposed SDP relaxation, including adaptive penalty updates and rigorous dual lower bounds. Section~\ref{sec:numerical_results} reports numerical results, and Section~\ref{sec:conclusion} concludes.

\subsection*{Notation}
For a finite set $\mathcal{O}$, $|\mathcal{O}|$ denotes its cardinality. Vector inequalities are interpreted element-wise. For matrices, $\langle A,B \rangle = \mathrm{trace}(A^\top B)$ denotes the Frobenius inner product, and $\|\cdot\|_F$ denotes the Frobenius norm. The set of symmetric $n\times n$ real matrices is $\mathcal{S}^n$, the positive semidefinite cone is $\mathcal{S}_+^n$, and the nonnegative orthant is $\mathbb{R}_+^n$. Trading periods are indexed by $t \in \mathcal{T}$ with a reference period $t_0$.

\section{The welfare-maximizing market-clearing model}
\label{sec:market_model}
We consider a day-ahead electricity market-clearing problem over a finite set of trading periods $\mathcal{T}$ following the policy in the Nordic energy market~\citep{CHATZIGIANNIS2016225}, and we omit the transmission constraints to focus on 1-area cases.

The market operator determines accepted quantities to maximize social welfare subject to market-balance constraints and order-type-specific feasibility requirements. The model includes three classes of orders: elementary hourly orders, block orders with logical constraints, and flexible hourly bids. Continuous variables are used for divisible orders, while binary variables represent activation decisions for nonconvex orders.

Let $\mathbf{x}_E$ denote the acceptance variables for elementary hourly orders, let $(\mathbf{x}_B,\mathbf{u}^{PB})$ denote the variables associated with block and profile-block orders, and let $\mathbf{x}_{FHB}$ denote the binary activation variables for flexible hourly bids. The welfare-maximization problem can then be written as the following MILP:
\begin{equation}\label{eq:mip_main}\tag{MILP}
    \begin{aligned}
        \max \; & \mathbf{c}_E^\top \mathbf{x}_E + \mathbf{c}_B^\top \mathbf{x}_B + \mathbf{c}_{FHB}^\top \mathbf{x}_{FHB} \\
        \text{s.t.}\; & \mathbf{A}_E \mathbf{x}_E + \mathbf{A}_B \mathbf{x}_B + \mathbf{A}_{FHB} \mathbf{x}_{FHB} = \mathbf{0}_{|\mathcal{T}|},\\
        & \mathbf{x}_E \in \mathcal{X}_E,\\
        & (\mathbf{x}_B,\mathbf{u}_{PB}) \in \mathcal{X}_B,\\
        & \mathbf{x}_{FHB} \in \mathcal{X}_{FHB}.
    \end{aligned}
\end{equation}
Here, the objective coefficients $\mathbf{c}_E$, $\mathbf{c}_B$, and $\mathbf{c}_{FHB}$ encode the welfare contribution of the three order classes, and the balance equation enforces market clearing in each trading period. The feasible sets $\mathcal{X}_E$, $\mathcal{X}_B$, and $\mathcal{X}_{FHB}$ capture the operational and logical requirements associated with the different order types.

For elementary hourly orders, the acceptance variables are continuous and bounded between zero and one. A subset of supply orders, denoted by $\mathcal{O}_{LG}$, is additionally subject to load-gradient constraints linking consecutive periods. Writing $q_{i,t}$ for the quantity associated with order $i$ at time $t$, these constraints define
\[
    \mathcal{X}_E :=
    \left\{
    \mathbf{x}_E \in [0,1]^{n_E \times |\mathcal{T}|} \;\middle|\;
    \begin{aligned}
    & q_{i,t} x_{i,t} - q_{i,t-1} x_{i,t-1} \le g_{up},\\
    & q_{i,t-1} x_{i,t-1} - q_{i,t} x_{i,t} \le g_{down},\\
    & \forall i \in \mathcal{O}_{LG},\ \forall t \in \mathcal{T}\setminus\{t_0\}
    \end{aligned}
    \right\}.
\]
These inequalities limit upward and downward changes in accepted quantities across adjacent periods.

The set $\mathcal{X}_B$ describes block and profile-block orders. These orders are nonconvex because binary activation decisions and logical relations govern their acceptance. We let $u_{PB}$ denote the relevant binary variables and write
\[
    \mathcal{X}_B :=
    \left\{
    (\mathbf{x}_B,\mathbf{u}_{PB}) \in [0,1]^{n_B} \times  \{0,1\}^{n_{PB}} \;\middle|\;
    \begin{aligned}
    & R u_{PB} \le x_B \le u_{PB},\\
    & \sum_{i \in \mathcal{O}_{eg}} x_i \le 1, \; \forall \mathcal{O}_{eg} \subseteq \mathcal{O}_B,\\
    & x_c \le \sum_{p \in P(c)} x_p, \; \forall c \in \mathcal{O}_L
    \end{aligned}
    \right\}.
\]
The inequalities $R u_{PB} \le x_B \le u_{PB}$ represent the profile-block structure, exclusivity is imposed over each exclusive group $\mathcal{O}_{eg}$, and the child--parent logic is modeled by the linkage constraints indexed by the set $\mathcal{O}_L$ of linked child orders, where $P(c)$ denotes the parent set associated with child order $c$.

Finally, flexible hourly bids are modeled using binary variables that indicate whether each bid is activated in each trading period. Since each flexible order can be accepted in at most one period, the feasible set is
\[
    \mathcal{X}_{FHB}
    :=
    \left\{
    \mathbf{x}_{FHB} \in \{0,1\}^{n_{FHB} \times |\mathcal{T}|}
    \;\middle|\;
    \sum_{t \in \mathcal{T}} x_{i,t} \le 1,\ \forall i \in \mathcal{O}_{FHB}
    \right\}.
\]

The model \eqref{eq:mip_main} provides a compact MILP representation of the welfare-maximizing market-clearing problem considered in this paper. This formulation will serve as the starting point for the lifted completely positive program reformulations and the tractable conic relaxations developed in the following sections.

\section{The equivalent completely positive reformulation}
\label{sec:CPP_reformulation}

Burer \cite{burer2009copositive} has proved that if a standard MIQP problem with only equations and nonnegativity constraints satisfies key assumptions that the bounds $[0,1]$ are induced by other constraints, then there exists an equivalent completely positive reformulation with standard conic relaxations techniques (squared equations and matrix lifting).  We can apply the same technique to form an equivalent CPP formulation for the market-clearing problem \eqref{eq:mip_main} and reformulate it by introducing slack variables $\mathbf{s}$ for all inequality constraints, including upper bounds $\mathbf{x} \leq 1$ for all variables.

For simplicity of notations in the standard reformulation \eqref{eq:milp_std}, let $\mathbf{x} := \left[ \mathbf{x}_E^\top, \mathbf{x}_B^\top, \mathbf{u}_{PB}^\top, \mathbf{x}_{FHB}^\top, \mathbf{s}^\top \right]^\top$ collect all original variables together with slack variables:
\begin{equation}\label{eq:milp_std}\tag{MILP-std}
    \begin{aligned}
    \min \; & -\mathbf{c}^\top \mathbf{x}\\
    \text{s.t.}~ & \mathbf{a}_t^\top \mathbf{x} = 0, \; \forall t \in \mathcal{T},\\
    & \mathbf{x} \in \mathcal{X}_{eq},\\
    & \mathbf{x} \in \mathbb{R}_+^{n_1} \times \{0,1\}^{n_2},
    \end{aligned}
\end{equation}
where $\mathbf{c}$ is the stacked objective vector, $\mathcal{X}_{eq}$ denotes the polyhedral set induced by the equality-form market constraints $\mathcal{X}_E$, $\mathcal{X}_B$, and $\mathcal{X}_{FHB}$, and the $\mathbb{R}_+^{n_1} \times \{0,1\}^{n_2}$ defines the joint space of the nonnegative variables and binary variables.

We first introduce the preliminary definitions of the completely positive program (CPP) before explaining the process of obtaining a CPP reformulation for \eqref{eq:milp_std}.

\begin{definition}[complete positive cone]
\label{def:CPP}
    The complete positive cone $\mathcal{C}^{*n}$ is the set of symmetric $n \times n$ real matrices: $$ \mathcal{C}^{*n} :=\{X \in \mathcal{S}^n \mid X= BB^\top, \exists~ B\in \mathbb{R}^{n\times m}_+\}.$$
\end{definition}
Note that a matrix in cone $\mathcal{C}^{*n}$ therefore admits a Gram representation using only nonnegative vectors. This property is the key device that allows binary and nonnegative continuous variables to be represented in a lifted conic model.

The standard technique to formulate a CPP relaxation is obtained by lifting the vector $x$ into a matrix variable $\mathbf{X} = \mathbf{x} \mathbf{x}^\top$. In the lifted space, quadratic products of variables are represented by entries of $\mathbf{X}$, while linear relations among components of $\mathbf{x}$ induce corresponding linear relations on $\mathbf{X}$. In particular, the augmented matrix
\[
    \bar{\mathbf{X}}:=
    \begin{pmatrix}
    1 & \mathbf{x}^\top\\
    \mathbf{x} & \mathbf{X}
    \end{pmatrix}
\]
is in the complete positive cone and we can relax $\mathbf{X} = \mathbf{x} \mathbf{x}^\top$ by $\mathbf{X} \in \mathcal{C}^{*n} $. Moreover, for all binary variables with index denoted as $\mathcal{O}_Z$,  we can also derive from $x^2 =x$, the diagonal consistency conditions $X_{ii}=x_i$ for all $i \in \mathcal{O}_Z$.  Another step to formulate the CPP is that each market-clearing equality $\mathbf{a}_t^\top \mathbf{x} = 0$ implies $\mathbf{a}_t^\top \mathbf{x} \mathbf{x}^\top \mathbf{a}_t = 0$, which, after lifting, becomes the linear constraint $\langle \mathbf{A}_t, \mathbf{X} \rangle = 0$, where $\mathbf{A}_t := \mathbf{a}_t \mathbf{a}_t^\top$ for each $t \in \mathcal{T}$. Likewise, the equality-form constraints defining $\mathcal{X}_{eq}$ generate a lifted feasible set for $\mathbf{X}$, which we denote by $\bar{\mathcal{X}}_{eq}$.

We thus obtain the following exact completely positive programming reformulation of \eqref{eq:milp_std}:
\begin{equation}\label{eq:cpp}\tag{CPP-eq}
    \begin{aligned}
        \min \; & -\mathbf{c}^\top \mathbf{x}\\
        \text{s.t.}\; & \mathbf{a}_t^\top \mathbf{x} = 0, \; \forall t \in \mathcal{T},\\
        & \langle \mathbf{A}_t, \mathbf{X} \rangle = 0, \; \forall t \in \mathcal{T},\\
        & \mathbf{x} \in \mathcal{X}_{eq},\; \mathbf{X} \in \bar{\mathcal{X}}_{eq},\\
        & \begin{pmatrix}
        1 & \mathbf{x}^\top\\
        \mathbf{x} & \mathbf{X}
        \end{pmatrix} \in \mathcal{C}^{*(1+n_1+n_2)},\\
        & X_{ii} = x_i, \; \forall i \in \mathcal{O}_Z,
    \end{aligned}
\end{equation}
where $\bar{\mathcal{X}}_{eq}$ denotes the collection of lifted linear constraints induced by the relations in $\mathcal{X}_{eq}$.

We omit the proof of the key assumption here, but it can be easily verified by Burer's theorem: \eqref{eq:cpp} is equivalent to the original MILP formulation. If $\mathbf{x}$ is feasible for \eqref{eq:milp_std}, then setting $\mathbf{X} = \mathbf{x} \mathbf{x}^\top$ yields a feasible point of \eqref{eq:cpp}, and both formulations have the same objective value. Conversely, any feasible solution of \eqref{eq:cpp} admits a completely positive lifted representation consistent with the binary diagonal conditions, which recovers a feasible solution of the original mixed-binary model.

Although formulation \eqref{eq:cpp} is exact and convex, optimization over the completely positive cone is computationally intractable in general. Its main value here is therefore structural: it reveals the market-clearing MILP as a lifted conic program and provides a natural foundation for relaxation. In the next section, we exploit this structure to derive relaxed and decomposed conic formulations, and then replace complete positivity by tractable doubly nonnegative approximations strengthened by problem-specific valid inequalities.

\section{The decomposed CPP and strengthened conic relaxation}
\label{sec:reformulation_decomposition_relaxation}
The exact CPP formulation \eqref{eq:cpp} provides a useful lifted representation of the market-clearing MILP, but it remains computationally intractable in general. In this section, we therefore develop more tractable conic models. We first introduce a relaxed lifted CPP formulation that avoids the explicit use of slack-variable lifting, and then exploit problem structure to derive a decomposed CPP model based on smaller local lifted matrices. We subsequently replace complete positivity by tractable semidefinite-based outer approximations and strengthen the resulting relaxations with DNN and problem-specific RLT inequalities.

\subsection{The decomposed CPP}
A first simplification is obtained by lifting only the original decision variables and retaining the inequality equations of the feasible region. Let $\mathcal{X}_{ineq}$ denote the polyhedral feasible set defined directly by the market-clearing constraints and the order-type inequalities, without introducing additional slack variables into the lifted vector. The resulting relaxed lifted CPP formulation is
\begin{equation}\label{eq:cpp_ineq}\tag{CPP-ineq}
    \begin{aligned}
        \min \; & -\mathbf{c}^\top \mathbf{x}\\
        \text{s.t.}\; & \mathbf{a}_t^\top \mathbf{x} = 0, \; \forall t \in \mathcal{T},\\
        & \langle \mathbf{A}_t, \mathbf{X} \rangle = 0, \; \forall t \in \mathcal{T},\\
        & \mathbf{x} \in \mathcal{X}_{ineq}, \; \mathbf{X} \in \bar{\mathcal{X}}_{ineq},\\
        & \begin{pmatrix}
        1 & \mathbf{x}^\top\\
        \mathbf{x} & \mathbf{X}
        \end{pmatrix} \in \mathcal{C}^{*(1+n_1+n_2)},\\
        & X_{ii} = x_i, \; \forall i \in \mathcal{O}_Z,\\
        & X_{ii} \le x_i, \; \forall i \in \mathcal{O}_R.
    \end{aligned}
\end{equation}
Here, $\mathcal{O}_Z$ and $\mathcal{O}_R$ denote the index sets of binary and continuous variables, respectively, and $\bar{\mathcal{X}}_{ineq}$ denotes the lifted relations induced by the inequalities in $\mathcal{X}_{ineq}$. Relative to \eqref{eq:cpp}, formulation \eqref{eq:cpp_ineq} reduces modeling overhead by avoiding an explicit lifted representation of slack variables, while preserving the basic conic lifting structure.

To further improve scalability, we exploit the sparsity induced by temporal coupling and order-type interactions.
Commonly used techniques, such as chordal decomposition, replace a large positive semidefinite constraint by a family of smaller clique-based constraints linked by consistency conditions, thereby reducing the memory and computational requirements of both first-order and second-order methods ~\cite{andersen2011chordal, zheng2020chordal}. Moreover, under certain sufficient conditions, the chordal-decomposed reformulation has the equivalent tightness as the original conic formulations \cite{grone1984positive, Drew01041998}.

Rather than imposing complete positivity on one global lifted matrix, we introduce several smaller lifted matrices associated with local variable groups. These groups are chosen to reflect the structure of the market-clearing model: period-wise balance constraints, load-gradient constraints linking consecutive periods, and couplings between block orders and flexible hourly bids. For each trading period $t \in \mathcal{T}$, let
\[
    \bar{\mathbf{X}}_t :=
    \begin{pmatrix}
    1 & \mathbf{x}_t^\top\\
    \mathbf{x}_t & \mathbf{X}_t
    \end{pmatrix}
    \in \mathcal{C}^{*(1+|\mathcal{O}_E|+|\mathcal{O}_B|+|\mathcal{O}_{FHB}|+|\mathcal{O}_{PB}|)},
\]
where $\mathbf{x}_t$ collects the variables in period $t$. For each $t \in \mathcal{T}\setminus\{t_0\}$, we also introduce a lifted matrix associated with the load-gradient coupling between periods $t-1$ and $t$:
\[
    \bar{\mathbf{X}}_{LG,t}:=
    \begin{pmatrix}
    1 & \mathbf{x}_{LG,t-1}^\top & \mathbf{x}_{LG,t}^\top & \mathbf{x}_B^\top & \mathbf{u}_{PB}^\top\\
    \mathbf{x}_{LG,t-1} & \multicolumn{4}{c}{\multirow{4}{*}{$\mathbf{X}_{LG,B,t}$}}\\
    \mathbf{x}_{LG,t} & \\
    \mathbf{x}_B & \\
    \mathbf{u}_{PB} &
    \end{pmatrix}
    \in \mathcal{C}^{*(1+2|\mathcal{O}_{LG}|+|\mathcal{O}_B|+|\mathcal{O}_{PB}|)},
\]
and, for each flexible hourly bid $i \in \mathcal{O}_{FHB}$, a lifted matrix capturing its interaction with block-order variables:
\[
    \bar{\mathbf{X}}_{B,FHB,i}:=
    \begin{pmatrix}
    1 & \mathbf{x}_{FHB,i}^\top & \mathbf{x}_B^\top & \mathbf{u}_{PB}^\top\\
    \mathbf{x}_{FHB,i} & \multicolumn{3}{c}{\multirow{3}{*}{$\mathbf{X}_{B,FHB,i}$}}\\
    \mathbf{x}_B & \\
    \mathbf{u}_{PB} &
    \end{pmatrix}
    \in \mathcal{C}^{*(1+|\mathcal{T}|+|\mathcal{O}_B|+|\mathcal{O}_{PB}|)}.
\]
The local lifted matrices overlap on shared variables and lifted entries. Consistency is enforced through linear projection operators that identify the corresponding submatrices of a global matrix $\mathbf{X}$ and the decomposed CPP formulation:
\begin{equation}\label{eq:cpp_ineq_decomp}\tag{CPP-ineq-decomposed}
    \begin{aligned}
        \min \; & -\mathbf{c}^\top \mathbf{x}\\
        \text{s.t.}\; & \mathbf{a}_t^\top \mathbf{x} = 0, \; \forall t \in \mathcal{T},\\
        & \langle \mathbf{A}_t, \mathbf{X} \rangle = 0, \; \forall t \in \mathcal{T},\\
        & \mathbf{x} \in \mathcal{X}_{ineq}, \; \mathbf{X} \in \bar{\mathcal{X}}_{ineq},\\
        & \mathcal{A}_t(\bar{\mathbf{X}}_t) = \mathbf{X}, \; \forall t \in \mathcal{T},\\
        & \mathcal{A}_{LG,t}(\bar{\mathbf{X}}_{LG,t}) = \mathbf{X}, \; \forall t \in \mathcal{T}\setminus\{t_0\},\\
        & \mathcal{A}_{B,FHB,i}(\bar{\mathbf{X}}_{B,FHB,i}) = \mathbf{X}, \; \forall i \in \mathcal{O}_{FHB},\\
        & \bar{\mathbf{X}}_t \in \mathcal{C}^*, \; \forall t \in \mathcal{T},\\
        & \bar{\mathbf{X}}_{LG,t} \in \mathcal{C}^*, \; \forall t \in \mathcal{T}\setminus\{t_0\},\\
        & \bar{\mathbf{X}}_{B,FHB,i} \in \mathcal{C}^*, \; \forall i \in \mathcal{O}_{FHB},
    \end{aligned}
\end{equation}
where $\mathcal{A}_{*}(\cdot)$ denote various linear  operators mapping between the decomposed submatrices and $\mathcal{X}$.

Formulation \eqref{eq:cpp_ineq_decomp} preserves the lifted conic logic of \eqref{eq:cpp_ineq}, but relaxes a single large cone constraint by a family of smaller cone constraints coupled through consistency relations. This decomposition is especially preferable computationally, since it prepares the model for sparse semidefinite relaxations and paralleled first-order splitting methods.

\subsection{The RLT strengthened DNN relaxations}

The complete positive cone in \eqref{eq:cpp}, \eqref{eq:cpp_ineq}, and \eqref{eq:cpp_ineq_decomp} are computationally intractable in general. We therefore relax those CP cones with tractable outer approximations. Again, we start with the definition of related conic cones.
\begin{definition}[positive semidefinite cone]
\label{def:PSD}
The positive semidefinite (PSD) cone $\mathcal{S}_+^n$ is the set of symmetric positive semidefinite matrices:
\[
    \mathcal{S}_+^n := \{\mathbf{X} \in \mathcal{S}^n \mid \mathbf{X} \succeq 0\}.
\]
\end{definition}
\begin{definition}[doubly nonnegative cone]
\label{def:DNN}
The doubly nonnegative (DNN) cone is the set of symmetric matrices that are both positive semidefinite and elementwise nonnegative:
\[
    \mathcal{DNN}^n := \mathcal{S}_+^n \cap \mathbb{R}_+^{n\times n}
    = \{\mathbf{X} \in \mathcal{S}^n \mid \mathbf{X} \succeq 0,\; \mathbf{X} \geq 0\}.
\]
\end{definition}
By the definition of the complete positive cone, it follows that $\mathcal{C}^{*n} \subseteq \mathcal{DNN}^n \subseteq \mathcal{S}_+^n$. This inclusion relation yields a hierarchy of tractable relaxations for the lifted formulations introduced above. The relaxation providing the lowest bounds for the minimization problem considered in this paper is obtained by retaining only positive semidefiniteness:
\begin{equation}\label{eq:sdp_relax}\tag{SDP}
    \mathbf{X} \succeq 0.
\end{equation}
We derive basic SDP relaxations of \eqref{eq:cpp}, \eqref{eq:cpp_ineq}, and \eqref{eq:cpp_ineq_decomp} by relaxing the CP cones to SDP cones. Although this relaxation captures convex quadratic structure, it does not enforce the element-wise nonnegativity that is intrinsic to complete positivity. A stronger approximation is obtained by imposing doubly nonnegative structure, i.e., relaxing the CP cones to DNN cones:
\begin{equation}\label{eq:dnn_relax}\tag{DNN}
    \mathbf{X} \in \mathcal{DNN}^n
    \; \text{or equivalently} \;
    \mathbf{X} \succeq 0,\;\; \mathbf{X} \geq 0.
\end{equation}
Replacing each complete positive constraint in \eqref{eq:cpp}, \eqref{eq:cpp_ineq}, and \eqref{eq:cpp_ineq_decomp} by the corresponding DNN constraint yields the DNN relaxations studied in this paper. In particular, when applied to the decomposed formulation \eqref{eq:cpp_ineq_decomp}, this replacement yields a structured primal semidefinite program with multiple smaller PSD blocks and element-wise nonnegativity constraints.

To further strengthen the DNN relaxation, we incorporate RLT inequalities. The basic idea is to multiply two valid linear inequalities and then replace the bilinear term $\mathbf{x}\mathbf{x}^\top$ by the lifted matrix $\mathbf{X}$. Specifically, suppose that $\mathbf{a}_1^\top \mathbf{x} - b_1 \geq 0$ and $\mathbf{a}_2^\top \mathbf{x} - b_2 \geq 0$. Then $(\mathbf{a}_1^\top \mathbf{x} - b_1)(\mathbf{a}_2^\top \mathbf{x} - b_2) \geq 0$, which expands to $\mathbf{a}_1^\top \mathbf{x}\mathbf{x}^\top \mathbf{a}_2 - (b_1 \mathbf{a}_2 + b_2 \mathbf{a}_1)^\top \mathbf{x} + b_1 b_2 \geq 0$. After lifting, this becomes a valid linear inequality:
\[
    \langle \mathbf{a}_1 \mathbf{a}_2^\top, \mathbf{X} \rangle
    - (b_1 \mathbf{a}_2 + b_2 \mathbf{a}_1)^\top \mathbf{x}
    + b_1 b_2 \geq 0.
\]
We use this construction to derive several families of valid inequalities adapted to the market-clearing structure. First, combining the bound constraints $0 \leq x_i \leq 1$ and $0 \leq x_j \leq 1$ yields the pairwise lifted bounds:
\begin{equation}\label{eq:rlt_lb_ub}
    1 - x_i - x_j \leq X_{ij} \leq \min\{x_i,x_j\}.
\end{equation}
These inequalities directly link lifted entries to the original decision variables, providing a simple yet effective strengthening of the DNN relaxation. Second, we derive RLT inequalities from the inner product of block-order constraints collected in $\mathcal{X}_B$:
\begin{equation}\label{eq:rlt_b}
    \mathbf{W}_B \mathbf{X}_B \leq \mathbf{f}_B.
\end{equation}
In addition, we generate RLT inequalities from products between block-order constraints and $0\leq x \leq 1$ constraints on the original variables:
\begin{equation}\label{eq:rlt_bx}
    \bar{\mathbf{W}}_B \mathbf{X}_t \leq \bar{\mathbf{f}}_B.
\end{equation}
Such inequalities are particularly useful in the decomposed setting, since they strengthen local lifted submatrices without requiring a fully dense global lifted matrix. Finally, consider the big-M constraints $r_i u_i \leq x_i \leq u_i, \; u_i^2 = u_i$. After lifting, these relations imply the valid inequalities 
\begin{equation}\label{eq:rlt_bigM}
    r_i u_i \leq X_{x_i,u_i} \leq u_i.
\end{equation}

These constraints tighten the interaction between profile-block activation variables and accepted quantities beyond what is imposed by the basic DNN relaxation alone.

In what follows, we refer to the model obtained by replacing complete positivity with \eqref{eq:sdp_relax} as the \texttt{SDP}, to the model obtained by replacing it with \eqref{eq:dnn_relax} as the \texttt{DNN}, and to the DNN relaxation strengthened with the RLT inequalities \eqref{eq:rlt_lb_ub}--\eqref{eq:rlt_bigM} together with the profile-block valid inequalities as the \texttt{DNN+RLT}. Among these, the \texttt{DNN+RLT} model provides the strongest tractable relaxation considered in this paper, while the decomposed version offers the most favorable balance between bound quality and computational scalability.

\section{The primal ADMM for the decomposed SDP relaxations}
\label{sec:ADMM_for_DNN}    
Applying the DNN relaxation to the decomposed lifted formulation \eqref{eq:cpp_ineq_decomp}, with optional RLT strengthening, yields a structured conic program with multiple local matrix variables coupled through linear consistency constraints. This structure is well-suited to operator-splitting methods. In this section, we develop an ADMM scheme for the resulting decomposed DNN and DNN+RLT relaxations, together with an adaptive penalty update and computable rigorous lower bounds for certified early termination. To describe the algorithmic framework, we write the decomposed DNN-type relaxations in the abstract form:
\begin{equation}\label{eq:cliqueSDP-P}\tag{decomDNN-P}
    \begin{aligned}
    \min \; & \sum_{i=1}^I \langle \mathbf{C}_i, \mathbf{X}_i \rangle\\
    \text{s.t.}\; & \sum_{i=1}^I \mathcal{A}_{ij}(\mathbf{X}_i) = \mathbf{Y}_j, \; \forall j \in J,\\
    & (\mathbf{X}_1,\dots,\mathbf{X}_I) \in \mathcal{X},\\
    & \mathbf{Y}_j \in \mathcal{DNN}^{n_j}, \; \forall j \in J.
    \end{aligned}
\end{equation}
Here, each matrix variable $\mathbf{X}_i$ corresponds to one local lifted block arising from the decomposition, the operators $\mathcal{A}_{ij}$ extract and assemble overlapping submatrices, and the set $\mathcal{X}$ collects all linear constraints, including the lifted balance equations, consistency relations, element-wise nonnegativity constraints, and, when applicable, the RLT inequalities introduced in Section~\ref{sec:reformulation_decomposition_relaxation}. The auxiliary variables $\mathbf{Y}_{j} $ represent the cone-constrained blocks on which positive semidefiniteness is enforced. Introducing dual multipliers $\mathbf{Z}_j$ for the coupling constraints $\sum_{i=1}^I \mathcal{A}_{ij}(\mathbf{X}_i) = \mathbf{Y}_j$, for all $j \in J$, the augmented Lagrangian associated with \eqref{eq:cliqueSDP-P} is
\begin{equation}\label{eq:aug_lag}
    \begin{aligned}
        \mathcal{L}_\rho(\{\mathbf{X}_i\},\{\mathbf{Y}_j\};\{\mathbf{Z}_j\}) = \sum_{i=1}^I \langle \mathbf{C}_i, \mathbf{X}_i \rangle + \sum_{j\in J} \left\langle \mathbf{Z}_j, \mathbf{Y}_j - \sum_{i=1}^I \mathcal{A}_{ij}(\mathbf{X}_i) \right\rangle \\
        + \frac{\rho}{2} \sum_{j\in J} \left\| \mathbf{Y}_j - \sum_{i=1}^I \mathcal{A}_{ij}(\mathbf{X}_i) \right\|_F^2,
    \end{aligned}
\end{equation}
where $\rho>0$ is the penalty parameter and $\|\cdot \|_F$ denotes the Frobenius norm . The corresponding ADMM iteration alternates between minimizing with respect to the primal block variables $\mathbf{X}_i$ and $\mathbf{Y}_j$ with other variables fixed, and updating the dual multipliers $\mathbf{Z}_i$. Given iterates $\bigl(\{\mathbf{X}_i^k\},\{\mathbf{Y}_j^k\},\{\mathbf{Z}_j^k\}\bigr)$, the ADMM update at iteration $k+1$ is defined by
\begin{subequations}\label{eq:admm_updates}
    \begin{align}
        (\mathbf{X}_i^{k+1})_{i=1}^I := & \argmin_{(\mathbf{X}_1,\dots,\mathbf{X}_I)\in\mathcal{X}} \left\{ \sum_{i=1}^I \langle \mathbf{C}_i, \mathbf{X}_i \rangle + \frac{\rho^k}{2} \sum_{j\in J} \left\| \mathbf{Y}_j^k - \sum_{i=1}^I \mathcal{A}_{ij}(\mathbf{X}_i) + \frac{1}{\rho^k}\mathbf{Z}_j^k \right\|_F^2 \right\},
        \label{eq:admm_x} \\
        \mathbf{Y}_j^{k+1} := & \operatorname{Proj}_{\mathcal{S}_+^{n_j}} \left( \sum_{i=1}^I \mathcal{A}_{ij}(\mathbf{X}_i^{k+1}) - \frac{1}{\rho^k}\mathbf{Z}_j^k \right), \; \forall j \in J,
        \label{eq:admm_y} \\
        \mathbf{Z}_j^{k+1} := & \mathbf{Z}_j^k + \rho^k \left( \mathbf{Y}_j^{k+1} - \sum_{i=1}^I \mathcal{A}_{ij}(\mathbf{X}_i^{k+1})
        \right), \; \forall j \in J,
        \label{eq:admm_z}
    \end{align}
\end{subequations}
where $\mathcal{A}^*$ are the adjoint operators of $\mathcal{A}$ and Proj$_{\succeq 0}$($\cdot$) denotes the projection onto the PSD cone.

The $\mathbf{X}$-update solves a convex quadratic program over the polyhedral set $\mathcal{X}$, which includes the linear equalities, element-wise nonnegativity constraints, and, when used, the RLT inequalities. The $\mathbf{Y}$-update consists of projections onto the positive semidefinite cones associated with the decomposed blocks. The $\mathbf{Z}$-update is the standard dual ascent step for the coupling constraints. Algorithm~\ref{alg:admm_cliquesdp} summarizes the resulting ADMM procedure.
\begin{algorithm}[htb!]
    \caption{ADMM for the decomposed DNN and DNN+RLT relaxations \eqref{eq:cliqueSDP-P}}
    \label{alg:admm_cliquesdp}
    \DontPrintSemicolon
    \KwIn{Penalty parameter $\rho^0 \in [\underline{\rho},\bar{\rho}]$, tolerance $\varepsilon>0$, and maximum iteration limit $K_{\max}$.}
    \KwInitialize{$(\mathbf{X}_i^0)_{i=1}^I$, $(\mathbf{Y}_j^0)_{j\in J}$, $(\mathbf{Z}_j^0)_{j\in J}$, and set $k \gets 0$}
    \While{$k < K_{\max}$}{
    $(\mathbf{X}_i^{k+1})_{i=1}^I \leftarrow \argmin_{(\mathbf{X}_1,\dots,\mathbf{X}_I)\in\mathcal{X}}
    \left\{
    \sum_{i=1}^I \langle \mathbf{C}_i,\mathbf{X}_i\rangle
    +\frac{\rho^k}{2}\sum_{j\in J}
    \left\|
    \mathbf{Y}_j^k-\sum_{i=1}^I\mathcal{A}_{ij}(\mathbf{X}_i)+\frac{1}{\rho^k}\mathbf{Z}_j^k
    \right\|_F^2
    \right\}$. \phantom{xxxxxxxxxxx} \Comment{$\mathbf{X}$-update}\;
    \For{$j\in J$}{
    $\mathbf{Y}_j^{k+1}\leftarrow
    \operatorname{Proj}_{\mathcal{S}_+^{n_j}}
    \left(
    \sum_{i=1}^I \mathcal{A}_{ij}(\mathbf{X}_i^{k+1})-\frac{1}{\rho^k}\mathbf{Z}_j^k
    \right)$.  \Comment{$\mathbf{Y}$-update}\;
    }
    \For{$j\in J$}{
    $\mathbf{Z}_j^{k+1}\leftarrow
    \mathbf{Z}_j^k+\rho^k\left(\mathbf{Y}_j^{k+1}-\sum_{i=1}^I\mathcal{A}_{ij}(\mathbf{X}_i^{k+1})\right)$.  \Comment{$\mathbf{Z}$-update}\;
    }
    Update $\rho^{k+1}$ according to \eqref{eq:adaptive_rho}.  \Comment{Stepsize update}\;
    Update $r^{k+1}$ and $s^{k+1}$ as \eqref{eq:residuals}\;
    \If{$\min\{r^{k+1}, s^{k+1} \} <\varepsilon$}{
    terminate.}\;
    $k \leftarrow k+1$.\;
    }
    \KwOut{$(\mathbf{X}_i^{k+1})_{i=1}^I$, $(\mathbf{Y}_j^{k+1})_{j\in J}$, $(\mathbf{Z}_j^{k+1})_{j\in J}$.}
\end{algorithm}
To monitor convergence, we use the standard ADMM primal and dual residuals:
\begin{equation}\label{eq:residuals}
    \begin{aligned}
        \mathbf{r}^{k+1} & := \left( \mathbf{Y}_j^{k+1} - \sum_{i=1}^I \mathcal{A}_{ij}(\mathbf{X}_i^{k+1}) \right)_{j\in J},\\
        \mathbf{s}^{k+1} & := \left( \rho^k \sum_{i=1}^I \mathcal{A}_{ij}(\mathbf{X}_i^{k+1}-\mathbf{X}_i^k) \right)_{j\in J}.
    \end{aligned}
\end{equation}
The primal residual measures violation of the coupling constraints, whereas the dual residual measures successive changes in the primal block variables through the coupling operators. As a first-order method, ADMM approaches struggle to reach a high precision as interior point methods, and the practical performance (e.g., convergent efficiency) of ADMM depends strongly on the choice of the penalty parameter. To speed up the convergence, we use the adaptive stepsize rule proposed by Lorenz and Tran-Dinh~\cite{lorenz2019non}:
\begin{equation}\label{eq:adaptive_rho}
    \begin{aligned}
        \rho^k & := (1-\omega^k)\rho^{k-1} + \omega^k \operatorname{Proj}_{[\underline{\rho},\bar{\rho}]} \left( \frac{\|\mathbf{u}^k\|_p}{\|\mathbf{y}^k\|_p} \right),\\
        \omega^k & := 2^{-k/1000},
    \end{aligned}
\end{equation}
where $[\underline{\rho},\bar{\rho}]$ is a prescribed interval of admissible penalty values, and the quantities $\mathbf{u}^k$ and $\mathbf{y}^k$ are the iteration-dependent vectors entering the adaptive rule. In our implementation, we use the $\ell_2$-norm. Moreover, to obtain a certificate for early termination, we compute a rigorous lower bound based on weak duality. Consider the Lagrangian associated with \eqref{eq:cliqueSDP-P}:
\[
    \mathcal{L}(\{\mathbf{X}_i\},\{\mathbf{Y}_j\};\{\mathbf{Z}_j\}) = \sum_{i=1}^I \langle \mathbf{C}_i, \mathbf{X}_i \rangle + \sum_{j\in J} \left\langle \mathbf{Z}_j,\, \mathbf{Y}_j-\sum_{i=1}^I \mathcal{A}_{ij}(\mathbf{X}_i) \right\rangle.
\]
The corresponding dual bound is obtained from
\begin{equation}\label{eq:dual_bound}
    \begin{aligned}
        \max_{\{\mathbf{Z}_j\}} \min_{\substack{(\mathbf{X}_1,\dots,\mathbf{X}_I)\in\mathcal{X}\\ \mathbf{Y}_j\in\mathcal{S}_+^{n_j},\, j\in J}} \mathcal{L}(\{\mathbf{X}_i\},\{\mathbf{Y}_j\};\{\mathbf{Z}_j\}).
    \end{aligned}
\end{equation}
Since the cone constraints are imposed on the variables $\mathbf{Y}_j$, minimizing over $\mathbf{Y}_j$ contributes the support function of the PSD cone, while minimizing over $\mathbf{X}$ produces a linear optimization problem over the polyhedral set $\mathcal{X}$. Thus, any dual-feasible collection of multipliers $\{\mathbf{Z}_j\}$ yields a valid lower bound on the optimal value of the decomposed DNN-type relaxation under consideration, including both the DNN and DNN+RLT models. Following the approach of~\cite{oliveira2018admm}, such bounds can be extracted from ADMM iterates with low precision and used to certify early stopping before full convergence is reached.

Taken together, the decomposed formulation, the DNN approximation, the optional RLT strengthening, the adaptive penalty update, and the rigorous lower-bound computation yield a practical first-order solution framework for the structured relaxations studied in this paper.

\section{Numerical results}
\label{sec:numerical_results}
This section reports numerical results for the proposed relaxations and first-order solution method. We study two questions: the bound quality of the lifted formulations under the SDP, DNN, and DNN+RLT relaxations, and the performance of ADMM on the decomposed DNN and DNN+RLT models. All experiments are conducted on synthetic instances that reflect the logical structure of day-ahead electricity market clearing, including elementary hourly bids, block orders, and flexible hourly bids. Interior-point SDP and DNN models are solved with MOSEK. The ADMM experiments are implemented in Python.

\subsection{Comparison of CPP formulations and relaxations}

The first experiments investigated the modeling power of conic relaxations for various CPP formulations and tested them on small-to-medium-sized instances. Those instances contain $|\mathcal{O}_E|=2$ elementary hourly orders, $|\mathcal{O}_{RB}|=4$ profile block orders, $|\mathcal{O}_{PB}|=4$ regular block orders, and $|\mathcal{O}_{FHB}|=4$ flexible hourly bids. The time horizon varies over $\{2,3,6,9,12,15,18,21,24\}$. Prices and quantities are generated synthetically. 
The optimal solutions and direct LP bounds are obtained by solving the MILP/LP problems in Gurobi, and Mosek solves the conic relaxations with precisions set as $10^{-6}$ for primal and dual feasibility tolerance and $10^{-4}$ for relative gap tolerance.

In Table~\ref{tab:ub_1}, we compare there level relaxations \texttt{sdp}, \texttt{dnn}, and \texttt{dnn+rlt} for the exact CPP formulation \eqref{eq:cpp}, the relaxed CPP formulation \eqref{eq:cpp_ineq}, and the decomposed CPP formulation \eqref{eq:cpp_ineq_decomp}(denoted as as \texttt{cpp\_eq}, \texttt{cpp\_ineq}, and \texttt{cpp\_ineq\_decomp} respectively). The optimal MILP solutions \texttt{opt.} and LP solutions \texttt{lp\_b} are given as reference, and the improvement of various conic relaxations relative to the LP lower bounds \texttt{lp\_b} is presented:
$$
    1-\frac{|obj_{\mathrm{conic}}-obj_{\mathrm{MILP}}|}{|obj_{\mathrm{LP}}-obj_{\mathrm{MILP}}|}.
$$

\begin{figure}[htp!]
\centering
\includegraphics[width=0.7\textwidth] {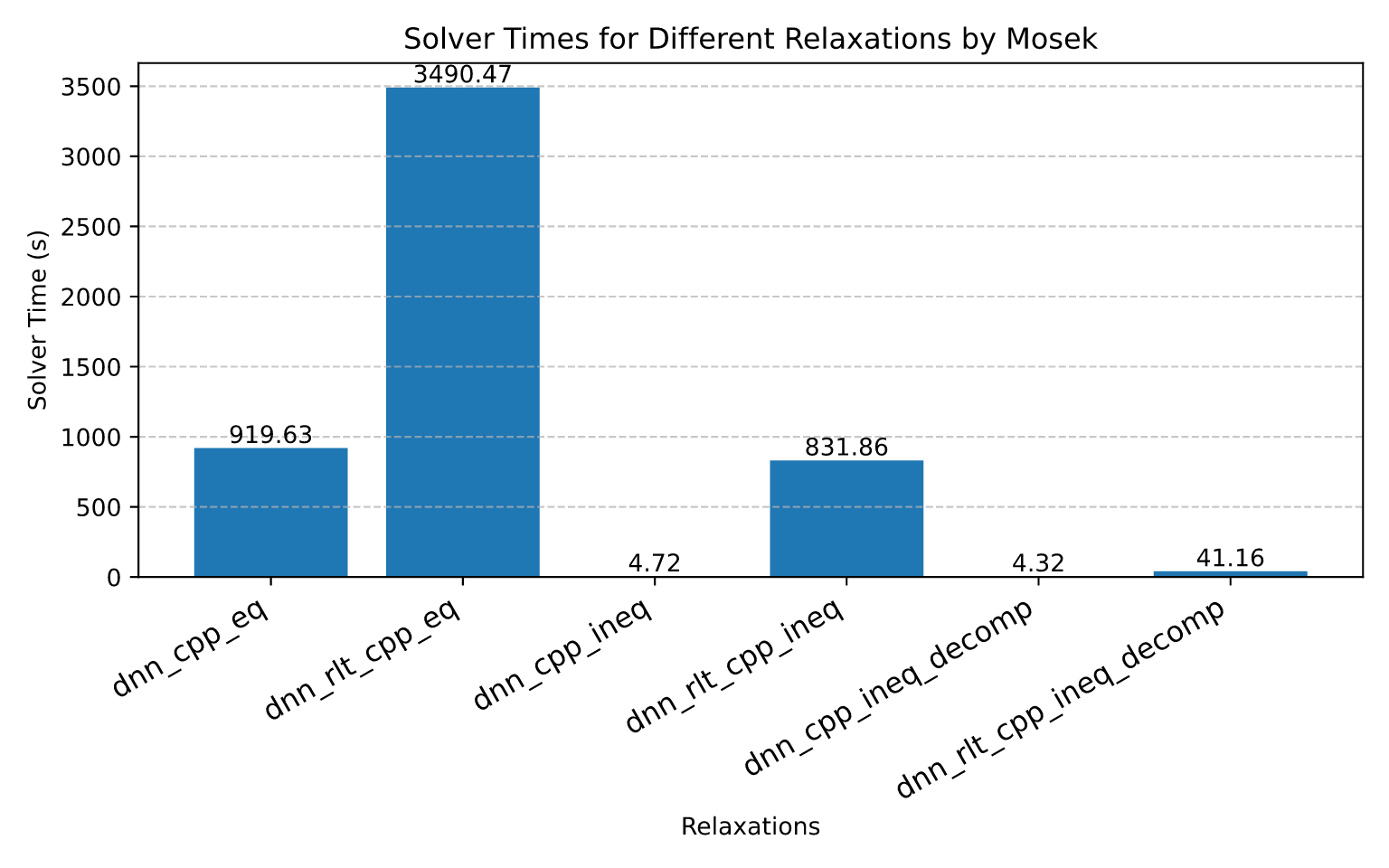}
\caption{The computation times (sec) of \texttt{dnn} and \texttt{dnn+rlt} by Mosek for $|\mathcal{T}| = 12$} \label{fig:computational_times}
\end{figure}

\begin{table}[htb!]
    \centering
    \resizebox{\textwidth}{!}{%
    \begin{tabular}{rrrrrrrrrrrr}
        \toprule
        \multirow{2}{*}{$|\mathcal{T}|$} & \multirow{2}{*}{\texttt{opt.}} & \multirow{2}{*}{\texttt{lp\_b}} & \multicolumn{3}{c}{\texttt{cpp\_eq}} & \multicolumn{3}{c}{\texttt{cpp\_ineq}} & \multicolumn{3}{c}{\texttt{cpp\_ineq\_decomp}}\\
        \cmidrule(lr){4-6} \cmidrule(lr){7-9} \cmidrule(l){10-12}
        & & & \texttt{sdp}& \texttt{dnn} & \texttt{dnn+rlt} & \texttt{sdp} & \texttt{dnn} & \texttt{dnn+rlt} & \texttt{sdp} & \texttt{dnn} & \texttt{dnn+rlt} \\
        \midrule
        2  & 150\,264   &   676\,582 & 0.69\% & 33.01\% & 99.78\% & 1.93\% & 19.95\% & 98.50\% & 1.89\% & 19.24\% & 98.51\% \\
        3  & 245\,008   &   940\,736 & 0.85\% & 34.17\% & 99.80\% & 2.59\% & 24.39\% & 97.83\% & 2.54\% & 22.98\% & 97.82\% \\
        6  & 402\,275   & 1\,799\,016 & 0.83\% & 43.33\% & 98.32\% & 2.74\% & 23.95\% & 97.76\% & 2.70\% & 22.85\% & 97.76\% \\
        9  & 456\,080   & 2\,287\,084 & 1.01\% & 55.59\% & 99.71\% & 3.12\% & 27.59\% & 99.26\% & 3.08\% & 26.42\% & 99.26\% \\
        12 & 555\,707   & 2\,897\,886 & 0.19\% & 21.48\% & 98.66\% & 0.66\% & 7.01\%  & 97.45\% & 0.62\% & 6.69\%  & 97.36\% \\
        15 & 656\,434   & 3\,543\,771 & 1.05\% & 65.68\% & --      & 3.38\% & 28.85\% & 99.88\% & 3.35\% & 27.60\% & 99.88\% \\
        18 & 776\,642   & 4\,231\,600 & 1.15\% & 68.84\% & --      & 3.50\% & 29.10\% & 99.90\% & 3.49\% & 27.91\% & 99.89\% \\
        21 & 896\,022   & 4\,871\,736 & 1.28\% & 77.29\% & --      & 3.50\% & 31.31\% & 99.91\% & 3.49\% & 29.66\% & 99.90\% \\
        24 & 1\,019\,784 & 5\,655\,408 & 0.25\% & --      & --      & 0.69\% & 7.61\%  & --      & 0.70\% & 7.56\%  & 99.59\% \\
        \bottomrule
    \end{tabular}
    }
    \caption{Improvement over the LP bound under a 10-hour time limit.}
    \label{tab:ub_1}
\end{table}
The exact CPP reformulation \eqref{eq:cpp} has the strongest modeling power compared to other formulations \eqref{eq:cpp_ineq} and \eqref{eq:cpp_ineq_decomp}: \texttt{cpp\_eq+dnn+rlt} has the best improvements, and similar performances are observed with \texttt{dnn} bounds for all formulations. However, it is worth mentioning that the rank of \texttt{cpp\_eq\_sdp}, \texttt{cpp\_ineq\_sdp}, and \texttt{cpp\_ineq\_decomp\_sdp} might be contradicting to the theoretical analysis, \texttt{cpp\_eq\_sdp} has the least improvements among all formulations due to the fact that the diagonal constraints $X_{ii}=x_i$ on the lifted matrices are not enforced while formulating the SDP relaxation on the CPP cone.  \texttt{cpp\_ineq\_decomp\_dnn+rlt} has achieved improvements only 1 point worse than the counterparts with a huge win on the computational efficiency: all the instances for \eqref{eq:cpp_ineq_decomp} are completed within the time limit in Table~\ref{tab:ub_1} while the strengthened conic relaxations \texttt{dnn} and \texttt{dnn + rlt } are computationally intractable when $|\mathcal{T}|\geq 12$; Figure~\ref{fig:computational_times} compares the computational times of \texttt{dnn} and \texttt{dnn+rlt} of all formulations, the reduction achieved by decomposed formulation is significant. 

This observation supports the decomposition strategy: it improves tractability with little loss in relaxation strength.
\subsection{The Primal ADMM for decomposed conic relaxations}
In this section, we focus on experiments involving larger-scale instances, using the two most efficient conic bounds: \texttt{cpp\_ineq\_decomp\_dnn} and \texttt{cpp\_ineq\_decomp\_dnn+rlt}.
We implement the ADMM update scheme Algorithm \ref{alg:admm_cliquesdp} to illustrate the computational efficiency of ADMM on those large-scale instances that Mosek fails to solve. The tolerance of the stopping criteria are $\epsilon_p =  10^{-3}$, $\epsilon_d = 10^{-4}$ and maximum iteration is $20000|\mathcal{T}|$. The instances set is divided into 5 groups based on the sizes (see Table~\ref{tab:instance_groups} denoted by $G_1$ -- $G_5$, with time horizons $|\mathcal{T}| \in \{4,8,12\}$). For each $(|\mathcal{T}|,G_{\cdot})$, Table~\ref{tab:admm_dnn_dnnrlt_results} reports summarized results for the ADMM dual bounds, the improvement over the LP bound, and the solution time (sec).
\begin{table}[htb!]
    \centering
    \begin{tabular}{cccccc}
        \toprule
        Group & $|\mathcal{O}_E|$ & $|\mathcal{O}_{RB}|$ & $|\mathcal{O}_{PB}|$ & $|\mathcal{O}_{FHB}|$ & Total orders for each time interval \\
        \midrule
        $G_1$ & 2 & 4 & 4 & 2 & 12 \\
        $G_2$ & 2 & 4 & 4 & 4 & 14 \\
        $G_3$ & 2 & 8 & 8 & 2 & 20 \\
        $G_4$ & 2 & 8 & 8 & 8 & 26 \\
        $G_5$ & 4 & 4 & 4 & 4 & 16 \\
        \bottomrule
    \end{tabular}
    \caption{Instance groups used in the ADMM experiments.}
    \label{tab:instance_groups}
\end{table}

The improvement over the LP bound is computed as
\[
    1-\frac{|obj_{\mathrm{ADMM}}-obj_{\mathrm{MILP}}|}{|obj_{\mathrm{LP}}-obj_{\mathrm{MILP}}|},
\]
and the higher values of which indicate stronger primal bounds returned by ADMM.

Table~\ref{tab:admm_dnn_dnnrlt_results} shows a clear advantage for the \texttt{dnn+rlt} bounds over the \texttt{dnn} bounds. Across all tested horizons and instance groups, \texttt{dnn+rlt} bounds achieve substantially larger improvements over the \texttt{lp} bounds. For $|\mathcal{T}|=4$, the average improvement ranges from 49.15\% to 72.06\% under  \texttt{dnn+rlt}, compared with 3.23\% to 21.02\% under DNN. For $|\mathcal{T}|=8$, the average \texttt{dnn+rlt} improvement lies between 33.43\% and 80.63\%, while the corresponding  \texttt{dnn} bounds remain between 1.67\% and 23.04\%. For $|\mathcal{T}|=12$, the differences becomes even more pronounced: \texttt{dnn+rlt} attains average improvements between 35.97\% and 92.46\%, whereas  \texttt{dnn} remains between 1.41\% and 28.25\%.
\begin{table}[htb!]
    \centering
    \resizebox{\textwidth}{!}{%
    \begin{tabular}{cccrrrrrrrr}
        \toprule
        \multirow{2}{*}{$|\mathcal{T}|$} & \multirow{2}{*}{Group} & \multirow{2}{*}{Statistic} & \multirow{2}{*}{\texttt{opt.}} & \multirow{2}{*}{\texttt{lp\_b}} & \multicolumn{2}{c}{ADMM dual bound} & \multicolumn{2}{c}{Improvement} & \multicolumn{2}{c}{Computational time (sec)} \\
        \cmidrule(lr){6-7} \cmidrule(lr){8-9} \cmidrule(l){10-11}
        & & & & & \texttt{dnn} & \texttt{dnn+rlt} & \texttt{dnn} & \texttt{dnn+rlt} & \texttt{dnn} & \texttt{dnn+rlt} \\
        \midrule

        \multirow{20}{*}{4}
        & \multirow{4}{*}{$G_1$} & Average   & 571\,188.34   & 1\,975\,759.17 & 1\,866\,048.79 & 880\,044.12   & 8.20\%  & 72.06\%  & 270.52  & 1\,090.69 \\
        &                       & Minimum   & 245\,292.00   & 1\,538\,506.60 & 1\,521\,937.01 & 361\,010.00   & 0.00\%  & 28.01\%  & 26.96   & 279.67   \\
        &                       & Maximum   & 1\,038\,940.99 & 2\,377\,674.75 & 2\,267\,757.68 & 1\,400\,429.21 & 26.14\% & 100.00\% & 466.94  & 2\,194.60 \\
        &                       & Std. Dev. & 307\,518.03   & 322\,080.00   & 336\,502.71   & 479\,120.77   & 10.26\% & 27.76\%  & 159.33  & 837.34   \\
        \cmidrule(lr){2-11}
        & \multirow{4}{*}{$G_2$} & Average   & 1\,018\,846.12 & 1\,968\,543.15 & 1\,841\,902.73 & 1\,441\,443.62 & 16.88\% & 49.19\% & 277.39 & 1\,225.15 \\
        &                       & Minimum   & 572\,674.00   & 1\,379\,381.71 & 887\,314.90   & 776\,450.67   & 0.00\%  & 11.79\% & 30.66  & 750.97   \\
        &                       & Maximum   & 1\,721\,754.45 & 2\,522\,181.10 & 2\,497\,412.89 & 1\,916\,639.40 & 71.56\% & 87.69\% & 674.30 & 1\,852.54 \\
        &                       & Std. Dev. & 448\,070.21   & 536\,811.38   & 692\,352.25   & 455\,199.52   & 30.79\% & 31.65\% & 266.28 & 402.60   \\
        \cmidrule(lr){2-11}
        & \multirow{4}{*}{$G_3$} & Average   & 3\,426\,854.18 & 5\,329\,690.11 & 5\,207\,919.46 & 3\,777\,861.39 & 5.61\%  & 69.89\% & 677.78  & 5\,244.84 \\
        &                       & Minimum   & 1\,424\,330.29 & 3\,375\,627.77 & 3\,234\,799.67 & 1\,595\,048.95 & 0.10\%  & 18.07\% & 68.39   & 870.19    \\
        &                       & Maximum   & 7\,558\,643.89 & 7\,712\,377.20 & 7\,712\,171.91 & 7\,684\,600.64 & 12.90\% & 99.83\% & 1\,048.03 & 7\,333.06 \\
        &                       & Std. Dev. & 2\,534\,936.29 & 1\,855\,924.47 & 1\,932\,341.44 & 2\,367\,057.92 & 5.49\%  & 33.29\% & 401.09  & 2\,737.20 \\
        \cmidrule(lr){2-11}
        & \multirow{4}{*}{$G_4$} & Average   & 5\,784\,187.64 & 6\,629\,556.33 & 6\,596\,811.52 & 6\,180\,807.88 & 3.23\%  & 49.15\% & 754.18  & 8\,971.96 \\
        &                       & Minimum   & 3\,933\,002.30 & 5\,145\,142.98 & 4\,987\,930.05 & 4\,668\,035.82 & 0.00\%  & 10.13\% & 206.15  & 4\,555.68 \\
        &                       & Maximum   & 8\,230\,979.66 & 8\,345\,518.60 & 8\,345\,511.43 & 8\,333\,913.83 & 13.98\% & 80.21\% & 1\,630.07 & 18\,171.26 \\
        &                       & Std. Dev. & 1\,876\,012.12 & 1\,249\,728.71 & 1\,298\,052.54 & 1\,550\,081.40 & 6.07\%  & 25.69\% & 581.04  & 5\,560.82 \\
        \cmidrule(lr){2-11}
        & \multirow{4}{*}{$G_5$} & Average   & 1\,885\,767.60 & 2\,681\,821.83 & 2\,491\,007.53 & 2\,205\,704.89 & 21.02\% & 61.52\% & 691.90 & 1\,240.07 \\
        &                       & Minimum   & 666\,232.80   & 1\,521\,135.00 & 1\,402\,664.30 & 1\,298\,573.42 & 5.97\%  & 26.03\% & 513.48 & 419.43   \\
        &                       & Maximum   & 3\,163\,162.77 & 3\,894\,304.40 & 3\,778\,390.08 & 3\,489\,174.69 & 58.96\% & 88.84\% & 878.68 & 1\,750.60 \\
        &                       & Std. Dev. & 890\,283.01   & 848\,809.40   & 860\,430.52   & 798\,321.69   & 21.53\% & 23.84\% & 160.13 & 582.69   \\
        \midrule

        \multirow{20}{*}{8}
        & \multirow{4}{*}{$G_1$} & Average   & 1\,486\,958.28 & 6\,818\,217.72 & 5\,528\,046.75 & 2\,908\,759.96 & 23.04\% & 73.79\% & 598.96  & 3\,210.10 \\
        &                       & Minimum   & 865\,783.41   & 4\,331\,066.86 & 3\,770\,727.18 & 1\,495\,249.76 & 9.75\%  & 28.76\% & 323.83  & 114.37   \\
        &                       & Maximum   & 2\,091\,416.00 & 8\,425\,628.50 & 7\,790\,610.98 & 6\,552\,139.13 & 51.25\% & 97.76\% & 694.01  & 5\,677.08 \\
        &                       & Std. Dev. & 513\,547.39   & 1\,725\,927.05 & 1\,696\,799.83 & 2\,098\,922.74 & 16.74\% & 28.65\% & 155.50  & 2\,686.14 \\
        \cmidrule(lr){2-11}
        & \multirow{4}{*}{$G_2$} & Average   & 1\,669\,072.55 & 6\,907\,138.99 & 6\,236\,345.96 & 3\,168\,055.33 & 13.10\% & 66.42\% & 800.93  & 4\,102.75 \\
        &                       & Minimum   & 824\,742.20   & 4\,805\,269.92 & 4\,079\,993.51 & 1\,471\,159.03 & 2.27\%  & 27.34\% & 516.32  & 2\,532.68 \\
        &                       & Maximum   & 2\,627\,844.35 & 8\,905\,242.50 & 7\,998\,347.97 & 4\,572\,534.35 & 21.44\% & 89.87\% & 1\,019.89 & 4\,803.99 \\
        &                       & Std. Dev. & 742\,891.65   & 1\,517\,204.36 & 1\,526\,582.14 & 1\,177\,070.88 & 6.89\%  & 25.11\% & 244.66  & 912.82   \\
        \cmidrule(lr){2-11}
        & \multirow{4}{*}{$G_3$} & Average   & 3\,612\,151.05 & 13\,314\,288.52 & 12\,677\,591.61 & 5\,712\,943.76 & 14.16\% & 80.63\% & 2\,319.22 & 18\,165.94 \\
        &                       & Minimum   & 1\,309\,013.22 & 10\,363\,740.37 & 8\,993\,891.97  & 2\,943\,075.08 & 0.01\%  & 60.92\% & 854.90   & 9\,212.38  \\
        &                       & Maximum   & 8\,917\,672.92 & 16\,937\,373.64 & 16\,909\,383.91 & 9\,479\,584.72 & 52.71\% & 95.53\% & 5\,296.30 & 25\,846.55 \\
        &                       & Std. Dev. & 3\,040\,909.30 & 2\,748\,638.13  & 3\,348\,730.20  & 3\,049\,635.73 & 22.84\% & 12.68\% & 1\,840.85 & 6\,577.09  \\
        \cmidrule(lr){2-11}
        & \multirow{4}{*}{$G_4$} & Average   & 10\,234\,246.48 & 16\,072\,826.79 & 16\,050\,137.98 & 14\,440\,906.69 & 1.67\%  & 33.43\% & 1\,353.47 & 42\,105.06 \\
        &                       & Minimum   & 4\,918\,338.93  & 12\,490\,586.49 & 12\,490\,740.32 & 8\,394\,300.43  & 0.00\% & 4.73\%  & 867.83   & 22\,999.65 \\
        &                       & Maximum   & 18\,206\,080.72 & 20\,928\,788.66 & 20\,859\,146.91 & 20\,179\,016.38 & 5.78\%  & 57.79\% & 2\,239.59 & 68\,278.97 \\
        &                       & Std. Dev. & 6\,119\,895.67  & 3\,551\,963.87  & 3\,528\,665.94  & 4\,915\,789.87  & 2.55\%  & 21.02\% & 526.52   & 16\,917.08 \\
        \cmidrule(lr){2-11}
        & \multirow{4}{*}{$G_5$} & Average   & 4\,588\,301.18 & 8\,399\,234.59 & 8\,229\,744.13 & 6\,416\,136.15 & 4.36\% & 50.28\% & 1\,079.50 & 5\,042.05 \\
        &                       & Minimum   & 3\,210\,013.16 & 6\,650\,392.32 & 6\,596\,881.40 & 3\,758\,439.83 & 0.01\% & 14.69\% & 116.76   & 393.55   \\
        &                       & Maximum   & 6\,207\,786.36 & 9\,912\,661.91 & 9\,770\,100.71 & 8\,430\,954.45 & 9.44\% & 89.62\% & 1\,846.91 & 7\,490.99 \\
        &                       & Std. Dev. & 1\,183\,816.01 & 1\,273\,803.29 & 1\,275\,634.32 & 1\,714\,151.33 & 3.36\% & 31.52\% & 641.67   & 2\,845.31 \\
        \midrule

        \multirow{20}{*}{12}
        & \multirow{4}{*}{$G_1$} & Average   & 1\,224\,396.54 & 8\,859\,546.41 & 6\,800\,126.13 & 1\,754\,992.62 & 28.25\% & 92.46\% & 1\,315.85 & 2\,279.54 \\
        &                       & Minimum   & 430\,956.00   & 5\,948\,843.91 & 4\,291\,000.22 & 698\,818.80   & 12.43\% & 83.63\% & 737.05   & 150.42    \\
        &                       & Maximum   & 2\,547\,355.72 & 12\,486\,133.83 & 11\,105\,056.81 & 3\,454\,364.80 & 45.43\% & 96.04\% & 1\,632.02 & 7\,745.34  \\
        &                       & Std. Dev. & 811\,358.06   & 2\,375\,174.27 & 2\,581\,881.21 & 1\,053\,879.32 & 11.93\% & 5.06\%  & 373.42   & 3\,194.59  \\
        \cmidrule(lr){2-11}
        & \multirow{4}{*}{$G_2$} & Average   & 2\,555\,826.73 & 11\,337\,552.83 & 10\,183\,165.77 & 3\,931\,426.38 & 11.80\% & 83.09\% & 1\,910.77 & 7\,837.36  \\
        &                       & Minimum   & 1\,348\,641.53 & 6\,224\,658.74  & 5\,807\,005.31  & 2\,298\,091.16 & 7.51\%  & 74.77\% & 1\,082.73 & 4\,837.39  \\
        &                       & Maximum   & 3\,870\,765.35 & 14\,943\,715.31 & 12\,656\,064.72 & 5\,593\,647.40 & 18.63\% & 90.01\% & 2\,466.94 & 10\,930.57 \\
        &                       & Std. Dev. & 1\,097\,641.12 & 3\,373\,390.75  & 2\,657\,972.79  & 1\,228\,464.71 & 4.91\%  & 5.73\%  & 538.78   & 2\,388.98  \\
        \cmidrule(lr){2-11}
        & \multirow{4}{*}{$G_3$} & Average   & 2\,468\,939.78 & 17\,692\,425.38 & 17\,227\,815.46 & 5\,713\,307.08 & 3.49\%  & 79.81\% & 3\,613.50 & 31\,148.28 \\
        &                       & Minimum   & 695\,042.83   & 9\,322\,266.45  & 8\,826\,959.27  & 2\,266\,640.40 & 0.00\% & 49.32\% & 689.30   & 9\,759.61  \\
        &                       & Maximum   & 3\,894\,014.43 & 21\,989\,602.05 & 21\,989\,702.21 & 12\,830\,432.83 & 7.34\%  & 94.21\% & 8\,108.90 & 76\,666.49 \\
        &                       & Std. Dev. & 1\,178\,390.06 & 5\,120\,157.85  & 5\,190\,559.41  & 4\,410\,865.40 & 3.65\%  & 19.71\% & 2\,882.05 & 26\,116.01 \\
        \cmidrule(lr){2-11}
        & \multirow{4}{*}{$G_4$} & Average   & 9\,566\,612.02 & 20\,659\,040.15 & 20\,558\,099.37 & 16\,226\,665.13 & 1.41\%  & 38.56\% & 3\,031.06 & 54\,627.62 \\
        &                       & Minimum   & 4\,269\,401.16 & 13\,372\,330.26 & 13\,368\,724.05 & 10\,353\,110.08 & 0.04\%  & 9.40\%  & 1\,882.96 & 39\,558.08 \\
        &                       & Maximum   & 17\,109\,206.98 & 25\,665\,786.31 & 25\,640\,881.42 & 24\,861\,345.76 & 6.52\%  & 59.26\% & 5\,002.05 & 75\,213.90 \\
        &                       & Std. Dev. & 5\,300\,792.04 & 4\,669\,764.31  & 4\,698\,567.53  & 5\,510\,888.57 & 2.86\%  & 19.47\% & 1\,289.72 & 15\,714.63 \\
        \cmidrule(lr){2-11}
        & \multirow{4}{*}{$G_5$} & Average   & 6\,531\,021.44 & 10\,643\,855.15 & 10\,440\,852.02 & 9\,419\,518.34 & 5.45\%  & 35.97\% & 1\,395.94 & 6\,753.45  \\
        &                       & Minimum   & 4\,449\,595.38 & 7\,829\,178.24  & 7\,120\,643.96  & 4\,990\,825.26 & 0.01\%  & 0.60\%  & 437.03   & 1\,535.74  \\
        &                       & Maximum   & 9\,569\,346.85 & 13\,001\,209.84 & 13\,000\,673.24 & 12\,964\,693.93 & 20.97\% & 83.99\% & 2\,418.01 & 10\,681.25 \\
        &                       & Std. Dev. & 2\,023\,588.85 & 1\,928\,260.92  & 2\,196\,313.88  & 3\,158\,091.99 & 8.84\%  & 40.16\% & 826.55   & 3\,368.84  \\
        \bottomrule
    \end{tabular}%
    }
    \caption{Dual bounds for the decomposed DNN and DNN+RLT relaxations obtained by Algorithm~\ref{alg:admm_cliquesdp}. ``Improvement'' is computed as $1-\frac{|obj_{\mathrm{ADMM}}-obj_{\mathrm{MILP}}|}{|obj_{\mathrm{LP}}-obj_{\mathrm{MILP}}|}$.}
    \label{tab:admm_dnn_dnnrlt_results}
\end{table}

The computational times show that the cost of attaining tighter bounds (e.g., \texttt{dnn+rlt}) is much higher. This difference is modest for the smaller instances, but becomes substantial as the horizon and order counts increase. For example, at $|\mathcal{T}|=12$, the average runtime for $G_4$ increases from 3\,031.06 seconds under DNN to 54\,627.62 seconds under \texttt{dnn+rlt}. A similar trend appears for the other larger groups.

Overall, Table~\ref{tab:admm_dnn_dnnrlt_results} shows that the proposed ADMM framework remains effective on both decomposed models. The decomposed DNN relaxation provides a useful baseline for fast first-order computation, while the decomposed DNN+RLT relaxation offers much stronger solutions at a higher computational cost. Together, these results support the use of ADMM as a practical solution method for the decomposed conic relaxations developed in this paper.

\section{Conclusion}
\label{sec:conclusion}
This paper developed CPP-based relaxations for day-ahead electricity market-clearing with nonconvex orders. Starting from a compact MILP formulation, we derived an equivalent completely positive reformulation, introduced relaxed and decomposed lifted variants, and constructed SDP, DNN, and DNN+RLT relaxations. The numerical results show that DNN yields a clear strengthening and that the addition of RLT inequalities substantially improves the bounds, often closing nearly the entire LP optimality gap. The decomposed formulation preserves essentially the same relaxation quality as the non-decomposed lifted model while significantly improving the computational tractability. For the decomposed models, we also proposed an efficient ADMM scheme with adaptive penalty updates and rigorous dual lower bounds, and the computational results indicate that it is effective on both the DNN and DNN+RLT relaxations. These results show that decomposed conic relaxations provide a promising and scalable approach to obtain strong bounds for nonconvex electricity market-clearing models.

\bibliographystyle{plain}
\bibliography{mybib}

\end{document}